# Mitigating Latent Mismatch in cVAE-Based Singing Voice Synthesis via Flow Matching

Minhyeok Yun and Yong-Hoon Choi, *Member, IEEE*

***Abstract*—Singing voice synthesis (SVS) aims to generate natural and expressive singing waveforms from symbolic musical scores. In cVAE-based SVS, however, a mismatch arises because the decoder is trained with latent representations inferred from target singing signals, while inference relies on latent representations predicted only from conditioning inputs. This discrepancy can weaken fine expressive acoustic details in the synthesized output. To mitigate this issue, we propose FM-Singer, a flow-matching-based latent refinement framework for cVAE-based singing voice synthesis. Rather than redesigning the acoustic decoder, the proposed method learns a continuous vector field that transports inference-time latent samples toward posterior-like latent representations through ODE-based integration before waveform generation. Because the refinement is performed in latent space, the method remains lightweight and compatible with a strong parallel synthesis backbone. Experimental results on Korean and Chinese singing datasets show that the proposed latent refinement improves objective metrics and perceptual quality while maintaining practical synthesis efficiency. These results suggest that reducing training-inference latent mismatch is a useful direction for improving expressive singing voice synthesis.**

**Code, pre-trained checkpoints, and audio demos are available at https://github.com/alsgur9368/FM-Singer.**

***Index Terms*—Singing voice synthesis, conditional variational autoencoder, flow matching, latent refinement, latent mismatch, expressive singing synthesis.**

## I. INTRODUCTION

SINGING voice synthesis (SVS) aims to generate natural and expressive singing waveforms from symbolic musical scores such as lyrics/phonemes, note pitch, and note durations. Compared to text-to-speech (TTS), singing voice synthesis must model a broader range of expressive phenomena—vibrato, timing offsets relative to the beat, dynamic accents, breathiness, and singer-specific timbral traits—while remaining faithful to strict musical constraints such as pitch targets and note boundaries. Although neural singing voice synthesis has substantially improved pitch accuracy and audio fidelity, generating fine-grained expressiveness remains challenging because these attributes are highly variable across singers and musical contexts and appear as subtle, localized deviations in pitch and spectral envelope.

A common strategy for the one-to-many nature of singing expression is to introduce latent variables that capture performance-specific variability beyond the score. End-to-end architectures derived from efficient TTS have been adapted to singing voice synthesis, where a conditional variational autoencoder (cVAE) latent variable is combined with adversarial learning to enable parallel generation and high-quality waveform synthesis [1]. VISinger and VISinger2 adopt a variational framework with adversarial training and signal-processing-inspired components, achieving strong results with efficient inference [2], [3]. Period Singer further highlights the importance of latent representations for singing characteristics by modeling periodic and aperiodic components with variational variants [4]. Despite these advances, cVAE-based singing voice synthesis typically uses a relatively simple score-conditioned prior and encourages prior–posterior alignment through Kullback–Leibler (KL) regularization. In practice, posterior latents inferred from real recordings during training can encode rich and multi-modal expressive cues, whereas inference uses samples from the prior; any residual mismatch can weaken fine expressive acoustic realization, including vibrato-like modulation and subtle timbral variation.

Recent advances in diffusion and flow-based generative modeling have been explored to improve detail and stability. Diffusion-based singing voice synthesis improves spectral fidelity via iterative denoising but can incur non-trivial inference cost due to multiple sampling steps [5]. In parallel, flow matching has emerged as a simulation-free method to train continuous normalizing flows by regressing a vector field along a chosen probability path, offering stable training and fewer numerical integration steps than many diffusion setups [6]. Flow matching has also been adopted for technique-controllable multilingual singing voice synthesis, indicating its potential for expressive generation. A representative example is TechSinger [7], which uses flow matching for technique-controllable singing voice synthesis. Consistency-model-style approaches reduce the number of steps while maintaining quality, including work targeting speech and singing synthesis [8].

This paper proposes FM-Singer, a flow-matching-based latent refinement module for cVAE-based singing voice synthesis. Rather than redesigning the acoustic decoder, the proposed method focuses on reducing the discrepancy

Date of submission March 13, 2026. This work was supported by Korea Institute for Advancement of Technology (KIAT) grant funded by the Korea Government (MOTIE) (RS-2024-00406796, HRD Program for Industrial Innovation); by the National Research Foundation of Korea (NRF) grant funded by the Korea government (MSIT) (RS-2025-16069933); and by the Research Grant of Kwangwoon University in 2024.

Corresponding author: Yong-Hoon Choi.

Minhyeok Yun and Yong-Hoon Choi are with the Fintech and AI Robotics (FAIR) Laboratory, the School of Robotics, Kwangwoon University, Nowon-gu, Seoul 01897, South Korea (e-mail: gurals9368@gmail.com; yhchoi@kw.ac.kr).

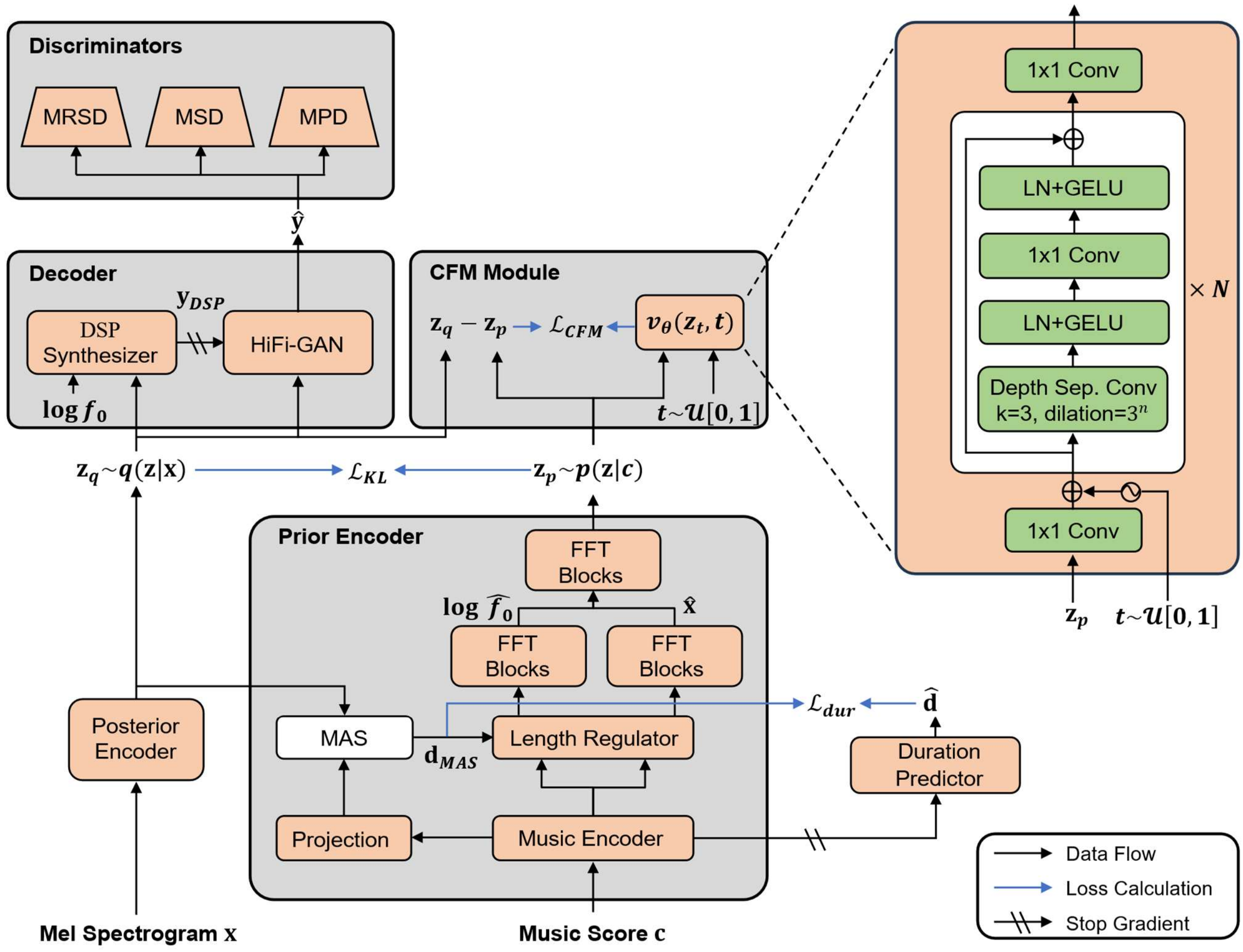

**Fig. 1.** **Overall training and inference pipeline of FM-Singer. The model learns a score-conditioned prior and a recording-conditioned posterior in a cVAE framework and refines inference-time prior samples using latent-space conditional flow matching before waveform generation.**

between the latent representations used during training and those available at inference. Specifically, FM-Singer refines condition-derived latent samples so that they better match the latent space observed by the decoder during training, thereby improving expressive acoustic realization while remaining lightweight and practical.

In summary, this work makes the following contributions. First, we highlight training-inference latent mismatch in cVAE-based singing voice synthesis as an important factor that can degrade fine expressive acoustic details. Second, we introduce a flow-matching-based latent refinement module that transports inference-time latent samples toward posterior-like latent representations. Third, we show through objective, perceptual, and efficiency evaluations that this refinement improves synthesis quality while preserving practical runtime performance.

In this work, we empirically examine training-inference latent mismatch in cVAE-based singing voice synthesis as a plausible source of degraded expressive acoustic realization. Our goal is not to claim a new flow-matching objective itself, but to demonstrate that flow matching can serve as an effective and lightweight latent refinement module within a cVAE-based SVS framework. The experimental results suggest that reducing this latent discrepancy helps improve the quality and expressiveness of synthesized singing voices.

The remainder of this paper is organized as follows. Section II presents the proposed architecture. Section III describes the training objective. Section IV reports experiments and analysis, and Section V concludes.

## II. Proposed Method

FM-Singer augments a cVAE-based singing voice synthesis backbone with a latent-space conditional flow matching (CFM) module. The overall training and inference pipeline is illustrated in Fig. 1. The model consists of a prior encoder, a posterior encoder, a latent refinement module trained by CFM, and a waveform generator trained with adversarial learning.

### A. Problem Formulation and Conditioning

Let $c$ denote the music-score conditioning, including phoneme/lyric tokens, note pitch, and note duration (or duration-related alignment). Let $y$ be the ground-truth singing waveform and $x = \mathrm{Mel}(y)$ be the corresponding mel-spectrogram. The goal is to synthesize waveform $\hat{y}$ that is faithful to $c$ while matching the expressive characteristics of real singing.

Expressive variability is modeled with latent variables $z$. During training, we learn a score-conditioned prior $p(z \mid c)$ and a recording-conditioned posterior $q(z \mid x)$. At inference time, only $c$ is available; therefore the model samples $z_p \sim p(z \mid c)$ and generates $\hat{y}$. A central issue in cVAE-based singing voice synthesis is that the decoder is trained using latent representations inferred from target singing signals, whereas at inference it must rely on latent representations predicted only from the conditioning inputs. This discrepancy can lead to degradation in fine expressive acoustic details. Note that FM-Singer does not modify the predicted fundamental frequence (F0) trajectory itself; instead, it refines latent acoustic representations so that the generated output more naturally realizes expressive phenomena such as vibrato-like modulation, micro-prosody, and subtle timbral variation.

### B. Prior and Posterior Encoders

The posterior encoder takes the mel-spectrogram $x$ and outputs the mean and variance of $q(z \mid x)$. The prior encoder takes music-score conditioning $c$ and outputs the parameters of $p(z \mid c)$. Both encoders are implemented using convolutional and residual blocks with conditioning mechanisms suitable for score-to-acoustic mapping. We implement both encoders using convolutional residual blocks inspired by WaveNet [9].

In practical singing voice synthesis setups, phoneme-level duration labels may be unavailable. Following the note-boundary-constrained alignment strategy introduced in Period Singer [4], we employ monotonic alignment search (MAS) constrained by note boundaries to estimate duration targets and supervise duration prediction. This design prevents cross-note alignment leakage, reduces timing ambiguity at note transitions, and stabilizes training by providing consistent duration targets.

### C. Latent Conditional Flow Matching

To explicitly reduce mismatch between $p(z \mid c)$ and $q(z \mid x)$, FM-Singer learns a conditional vector field that transports a prior latent sample toward a posterior latent sample. The process is summarized in Fig. 1.

Let $z_p \sim p(z \mid c)$ and $z_q \sim q(z \mid x)$. We sample $t \sim \mathcal{U}[0,1]$ and define a straight-line interpolation following flow-matching training [6]:

$$z_t = (1 - t)z_p + tz_q. \tag{1}$$

The target velocity along this path is:

$$u_t = \frac{dz_t}{dt} = z_q - z_p. \tag{2}$$

We train a neural vector field $v_\theta$ to match the target velocity:

$$\mathcal{L}_{CFM} = \mathbb{E}_{t,\mathbf{z}_p,\mathbf{z}_q}[\|v_\theta(z_t, t) - u_t\|_2^2]. \tag{3}$$

The vector field $v_\theta$ takes the interpolated latent $z_t$ and continuous time $t$ as inputs, where $t$ is encoded using a sinusoidal or learned time embedding and injected into the residual blocks. In practice, score-side conditioning can be provided implicitly via the endpoint sampling (through $z_p$) and/or explicitly by concatenating a compact conditioning projection to the input of $v_\theta$. The role of the proposed flow-matching module is not to replace the cVAE decoder, but to refine inference-time latent samples so that they move toward posterior-like regions of the latent space. In this way, the decoder receives latent representations that are more consistent with those observed during training.

At inference time, we sample $z_p \sim p(z \mid c)$ and solve the following ordinary differential equation (ODE):

$$\frac{dz}{dt} = v_\theta(z, t), \qquad z(0) = z_p, \tag{4}$$

to obtain a refined latent $z(1)$, denoted by $\hat{z}$. Numerical integration is implemented using torchdiffeq [10] with a Dormand–Prince (DOPRI5) solver [11]. This refinement step is lightweight because it operates in latent space, and it is applied once per utterance (or per segment), after which the refined latent is consumed by the waveform generator.

We apply the refinement either once per utterance or per fixed-length segment depending on the training/inference setup; segment-wise refinement can improve stability for long recordings while keeping memory usage bounded. The refinement is performed only in latent space, so its computational cost is typically limited compared with waveform generation. Importantly, the ODE solution can be interpreted as a learned continuous transport that reduces the gap between inference-time prior samples and training-time posterior latents.

### D. CFM Module and ODE Settings

The vector field estimator $v_\theta$ is implemented as a compact convolutional residual stack. Specifically, we use a hidden dimension of 192 with kernel size 3 and stack four dilated depth-separable convolution (DDSConv) blocks. The dilation is increased geometrically to expand the receptive field (e.g., 3, 5, 7, 9), and dropout with probability $p = 0.1$ is applied within the DDSConv blocks for regularization. These choices provide sufficient modeling capacity for latent transport while keeping the CFM module lightweight relative to the generator.

We use dilations to enlarge the receptive field without increasing parameter count, allowing $v_\theta$ to model both short-range and longer-range temporal correlations in the latent trajectory. This is useful for capturing fine temporal expressive variations, including vibrato realization and micro-prosody, while preserving overall latent coherence during refinement. The lightweight design keeps the refinement module small enough to be attached to an existing cVAE backbone without noticeably affecting training stability.

**TABLE 1. Hyperparameter settings for the FM-Singer generator, latent refinement module, and ODE solver**

| Layer | Hyperparameters | Values |
|---|---|---|
| CFM Module | Hidden channel | 192 |
| | Number of DDSConv blocks | 4 |
| | DDSConv dilation rates | [3,5,7,9] |
| | Kernel size | 3 |
| | Dropout | 0.1 |
| | ODE solver | DOPRI5 |
| | Tolerances | $1 \times 10^{-5}$ |
| | Max step | 0.1 |
| Prior Encoder | Number of hidden channels | 256 |
| | Number of FFT filter channels | 1024 |
| | Number of FFT blocks | 4 |
| Posterior Encoder | Number of hidden channels | 192 |
| | WaveNet kernel size | 5 |
| | Number of WaveNet blocks | 16 |
| Decoder | Number of hidden channels | 192 |
| | Upsampling rates | [8,8,4,2] |
| | Upsampling kernel sizes | [16,16,8,4] |

For inference-time ODE integration, we set both absolute and relative tolerances to $1 \times 10^{-5}$ and cap the maximum step size at 0.1, which enforces at least 10 integration steps over $t \in [0,1]$ even when the learned dynamics are smooth. Hyperparameters are summarized in Table 1.

### *E. Generator and Discriminators*

The waveform generator follows a generative adversarial network (GAN)-based design. As shown in Fig. 1, the generator converts the refined latent $\hat{z}$ and pitch-related conditions into waveform output $\hat{y}$. To train the generator, we employ three discriminators:

$$\mathcal{D} = \{D_{\mathrm{MRSD}}, D_{\mathrm{MPD}}, D_{\mathrm{MSD}}\}, \tag{5}$$

where $D_{\mathrm{MRSD}}$ is a multi-resolution spectrogram discriminator (MRSD) [12], $D_{\mathrm{MPD}}$ is a multi-period discriminator (MPD), and $D_{\mathrm{MSD}}$ is a multi-scale discriminator (MSD). These discriminators provide complementary supervision: MPD is effective at modeling periodic structures, MSD captures multi-scale time-domain realism, and MRSD constrains spectro-temporal realism across multiple time–frequency resolutions.

This choice follows common GAN vocoder practice, where multi-period and multi-scale discriminators improve periodicity and multi-resolution realism in the time domain [13], and spectrogram-based discriminators encourage consistent time–frequency structure at multiple resolutions. Using all three discriminators provides more reliable gradients across diverse singing conditions, including sustained vowels, rapid note changes, and high-pitch regions where artifacts are more likely to appear.

We further adopt feature matching and mel-spectrogram reconstruction losses to stabilize adversarial learning and to improve perceptual quality.

## III. Training Objective Details

FM-Singer is optimized by combining cVAE regularization, latent CFM loss, and GAN-based waveform generation losses, along with auxiliary terms. The training objective is designed to (i) align the inference-time prior with the training-time posterior, (ii) synthesize high-fidelity waveforms, and (iii) preserve accurate timing and pitch realization.

### *A. KL Regularization for the cVAE*

We regularize the latent space by minimizing the KL divergence between posterior and prior:

$$\mathcal{L}_{KL} = \mathrm{KL}\big(q(z \mid x) \,\|\, p(z \mid c)\big). \tag{6}$$

This term encourages the score-conditioned prior $p(z \mid c)$ to match the recording-conditioned posterior $q(z \mid x)$, reducing the discrepancy between training-time and inference-time latent usage. In our setting, KL regularization alone is often insufficient to fully align expressive, multi-modal posterior latents, motivating the additional latent transport term in (3).

### *B. Generator Losses*

Let $y$ and $\hat{y}$ denote the ground-truth and generated waveforms, respectively. Using the discriminators in (5), we adopt a least-squares adversarial objective for the generator:

$$\mathcal{L}_{\mathrm{adv}}(G) = \sum_{D_k \in \mathcal{D}} \mathbb{E}_{\hat{y}}\left[(D_k(\hat{y}) - 1)^2\right], \tag{7}$$

where the expectation is approximated by the mini-batch average of $\hat{y}$ generated from paired $(c, x)$ samples.

To stabilize adversarial training and encourage perceptual similarity, we use a feature matching loss. Let $D_k^{(\ell)}(\cdot)$ denote the activation at the $\ell$-th layer of discriminator $D_k$, and let $N_{k,\ell}$ be the number of elements in that activation. The feature matching loss is

$$\mathcal{L}_{\mathrm{FM}} = \sum_{D_k \in \mathcal{D}} \sum_{\ell} \frac{1}{N_{k,\ell}} \left\| D_k^{(\ell)}(y) - D_k^{(\ell)}(\hat{y}) \right\|_1. \tag{8}$$

The feature matching loss is computed over intermediate discriminator layers (typically all layers except the final output layer), which encourages the generator to match multi-level representations of real audio. Normalizing by $N_{k,\ell}$ prevents layers with larger activations from dominating the objective and improves training balance across discriminators. This term is especially important for singing voice synthesis because it reduces over-sharpened artifacts while preserving harmonic structure and fine temporal expressive variations.

We additionally apply a mel-spectrogram reconstruction loss:

$$\mathcal{L}_{\text{mel}} = \|\text{Mel}(y) - \text{Mel}(\hat{y})\|_1. \tag{9}$$

Here, $\text{Mel}(\cdot)$ denotes a fixed mel-spectrogram transform with the same analysis parameters used to generate the training targets $x$. The mel reconstruction term provides a strong signal for spectral envelope and overall intelligibility, complementing discriminator feedback which may focus on finer time-domain realism.

The GAN-related generator loss is

$$\mathcal{L}_G = \mathcal{L}_{\text{adv}}(G) + \lambda_{\text{FM}}\mathcal{L}_{\text{FM}} + \lambda_{\text{mel}}\mathcal{L}_{\text{mel}}, \tag{10}$$

where $\lambda_{FM} = 2$ and $\lambda_{mel} = 45$, following VISinger2 [3].

### C. Additional Generator Losses

We use MAS-based duration estimates $d_{\text{MAS}}$ as targets for the predicted duration $d_{\text{pred}}$:

$$\mathcal{L}_{\text{dur}} = \left\|d_{\text{MAS}} - d_{\text{pred}}\right\|_2^2. \tag{11}$$

Let $y_{\text{DSP}}$ denote the waveform produced by a DSP synthesizer. We define a DSP loss as

$$\mathcal{L}_{\text{DSP}} = \lambda_{\text{DSP}}\|\text{Mel}(y_{\text{DSP}}) - \text{Mel}(y)\|_1, \tag{12}$$

where $\lambda_{\text{DSP}} = 45$. This term encourages the DSP branch to remain consistent with the target and supports stable training when the generator leverages DSP-guided components. The DSP-based supervision provides an additional anchor for spectral consistency, which can improve robustness when the generator is still learning stable waveform synthesis. It also helps prevent pitch-related collapse in difficult regions by encouraging the generated content to remain close to a signal-processing-guided reference in the mel domain.

We further regularize the prior encoder using an auxiliary prediction loss on continuous pitch and mel-spectrogram. Let $\widehat{\log f_0}$ and $\hat{x}$ be the predicted $\log f_0$ and mel-spectrogram, respectively. Then

$$\mathcal{L}_{aux} = \left\|\log f_0 - \widehat{\log f_0}\right\|_2^2 + \|x - \hat{x}\|_1. \tag{13}$$

The auxiliary predictions act as regularizers for the prior side, encouraging the score-conditioned pathway to encode pitch-relevant and spectral cues that are useful at inference time. This is particularly beneficial because the prior encoder must provide informative latents without access to the target recording, and the auxiliary losses help reduce under-conditioning in pitch-sensitive singing segments.

### D. Discriminator Loss

Each discriminator $D_k \in \mathcal{D}$ is optimized using the least-squares objective:

$$\mathcal{L}_{\text{adv}}^{D_k} = \mathbb{E}_y[(D_k(y) - 1)^2] + \mathbb{E}_{\hat{y}}[D_k(\hat{y})^2]. \tag{14}$$

The total discriminator loss is

$$\mathcal{L}(D) = \sum_{D_k \in \mathcal{D}} \mathcal{L}_{\text{adv}}^{D_k}. \tag{15}$$

### E. Final Objective

The generator is optimized with

$$\mathcal{L}(G) = \mathcal{L}_G + \mathcal{L}_{\text{KL}} + \mathcal{L}_{\text{DSP}} + \mathcal{L}_{\text{dur}} + \mathcal{L}_{aux} + \lambda_{\text{CFM}}\mathcal{L}_{\text{CFM}}. \tag{16}$$

We alternately update the generator and discriminators by minimizing $\mathcal{L}(G)$ and $\mathcal{L}(D)$, respectively. We set $\lambda_{\text{CFM}} = 1$ in all experiments unless otherwise stated. During training, we alternately update the discriminators using (14)–(15) and the generator using (16) with the same batch of paired $(c, x)$ samples, following standard GAN training practice. This joint objective ensures that the latent space is regularized (KL), transported toward expressive posteriors (CFM), and decoded into high-fidelity waveforms (GAN and reconstruction losses).

## IV. Experiments

To comprehensively assess the effectiveness of the proposed latent transport, we design experiments to answer two questions: (i) whether latent-space conditional flow matching improves fine-grained expressiveness while preserving the efficiency of a parallel cVAE backbone, and (ii) whether the benefits generalize across different languages and datasets. We therefore evaluate FM-Singer on both Korean and Chinese benchmarks, and compare against strong cVAE-based baselines and representative refinement strategies. Our novelty does not lie in introducing a new flow-matching objective, but rather in formulating latent transport as a plug-and-play bridge between the conditional prior and posterior within a parallel cVAE SVS backbone, and evaluating its effect on synthesis quality and expressive acoustic realization.

Our evaluation protocol includes both objective and perceptual measurements. Objective metrics quantify spectral and pitch fidelity, while subjective listening tests reflect perceptual naturalness and expressiveness. We further provide qualitative visualizations to highlight how latent refinement affects time–frequency structure and fine temporal expressive variations, including vibrato-like patterns. Unless otherwise stated, we keep backbone configurations consistent across systems for a fair comparison, and summarize key training and inference hyperparameters in Table 1. The main quantitative results are reported in Tables 2 and 3 (with MOS reported for the Korean dataset), and qualitative examples are shown in Fig. 2.

**TABLE 2. Results on the Korean singing voice dataset after 70k training steps. We report MCD, F0 RMSE, and MOS (95% confidence interval).**

| Model | MCD ↓ | F0 RMSE ↓ | MOS ↑ |
|---|---|---|---|
| Ground Truth | - | - | 4.592 (± 0.05) |
| VISinger2 [3] | 6.328 | 39.4 | 3.347 (±0.07) |
| VISinger2 NF | 5.784 | 39.1 | 3.569 (±0.07) |
| FM-Singer (ours) | **4.815** | **35.8** | **4.039 (±0.06)** |

**TABLE 3. Results on the Chinese singing voice dataset after 500k training steps. We report MCD and F0 RMSE.**

| Model | MCD ↓ | F0 RMSE ↓ |
|---|---|---|
| Ground Truth | - | - |
| VISinger2 [3] | 3.587 | 26.7 |
| VISinger2 NF | 2.939 | 25.5 |
| FM-Singer (ours) | **2.703** | **25.2** |

### *A. Dataset*

We evaluate FM-Singer on two benchmarks. The first is a Korean singing dataset consisting of studio-quality recordings paired with score information, where phoneme-level duration labels may be missing and note-boundary MAS is used for duration supervision. The second is a Chinese singing benchmark based on OpenCpop [14], a publicly available corpus designed for singing voice synthesis research.

### *B. Baselines*

We compare FM-Singer with VISinger2 [3] and a variant without latent flow refinement (VISinger2 NF) to isolate the effect of latent transport. We also evaluate two-stage refinement pipelines based on a duration/pitch-aware acoustic model and a neural vocoder, following FastSpeech-style modeling [15] with GAN-based refinement/vocoding such as RefineGAN [16]. This comparison helps to contextualize the trade-offs between various refinement strategies and efficient parallel waveform generation.

We note that TechSinger [7], a recent flow-matching-based technique-controllable SVS system, is not included in our quantitative tables. This is because its released setup requires additional technique-related conditioning and an associated pipeline that cannot be matched to our current datasets and evaluation protocol without substantial re-engineering; therefore, we restrict comparisons to baselines that can be reproduced under matched conditions in our pipeline. We emphasize that TechSinger targets technique-controllable SVS, whereas our focus is bridging the cVAE prior–posterior gap via latent transport while preserving a parallel backbone.

### *C. Implementation Details*

We keep the backbone architecture close to VISinger2 [3] to isolate the effect of latent refinement. The CFM module uses a hidden dimension of 192, kernel size 3, and four DDSConv blocks with dilation increasing geometrically, with dropout probability 0.1. For inference-time ODE integration, we use DOPRI5 with absolute and relative tolerances $1 \times 10^{-5}$ and maximum step size 0.1. The hyperparameter configuration is summarized in Table 1. Models are trained on a single NVIDIA A100 (80GB) GPU.

### *D. Evaluation Metrics*

We report mel-cepstral distortion (MCD) as a spectral distance metric [17] and fundamental frequency (F0) root mean square error (RMSE) to quantify pitch trajectory error. MCD is computed on aligned sequences using standard mel-cepstral analysis. F0 RMSE is computed on voiced regions using a continuous log f0 representation with appropriate handling of unvoiced segments. For perceptual quality, we conduct a mean opinion score (MOS) test on the Korean dataset using a five-point scale. Due to evaluation-budget constraints, subjective listening tests were conducted on the Korean dataset only, while the Chinese dataset was used for objective evaluation. This setup allows us to assess both perceptual quality and cross-dataset generalization without changing the overall experimental protocol. Participants were instructed to use headphones during the test. MOS is reported on a 1–5 scale with 95% confidence intervals.

### *E. Quantitative Results*

Table 2 reports objective and subjective results on the Korean dataset after 70k training steps. FM-Singer improves MOS while reducing MCD and F0 RMSE compared with VISinger2 and VISinger2 NF, suggesting that latent refinement improves synthesis quality while preserving overall naturalness. The comparison with VISinger2 NF isolates the effect of removing latent transport while keeping the backbone configuration the same. The additional gain of FM-Singer over VISinger2 NF suggests that transporting condition-derived latent samples toward posterior-like latent representations, rather than merely introducing an extra refinement stage, is an important factor in the observed improvement in expressive synthesis quality.

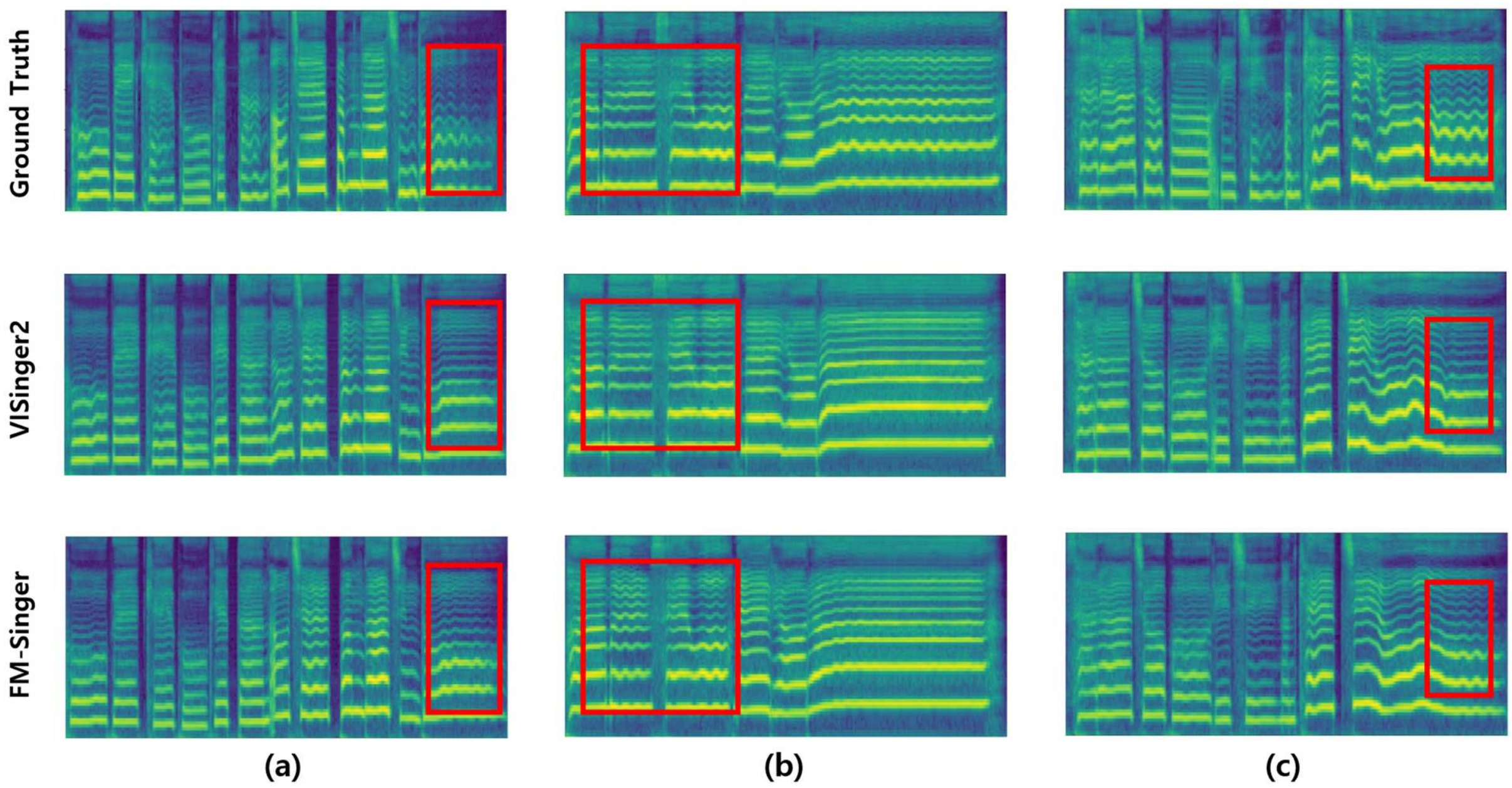

**Fig. 2.** **Qualitative comparison of synthesized singing samples for the same lyric segment. (a) Mel-spectrograms, (b) waveforms, and (c) F0 contours. Compared with VISinger2 and VISinger2 NF, FM-Singer better preserves harmonic structure and fine temporal expressive variations while producing outputs that are closer to the ground-truth recording. The red boxes indicate the same regions across all systems, selected from the ground-truth recording where fine expressive temporal variations are most prominent.**

**TABLE 4.** **Distance comparison between posterior latent samples and condition-derived latent samples before and after latent refinement. We report the mean, median, and 90th percentile (P90) of the latent distance distribution. The relative improvement is computed as** $\mathrm{Rel.} = 100 \times \left(1 - \|\hat{z} - z_q\|_2 / \|z_p - z_q\|_2\right)\%$. **A smaller distance and a larger Rel. indicate that the refined latent representation is closer to the posterior latent space observed during training.**

| Model | $\|z_p - z_q\|_2 \downarrow$ | $\|\hat{z} - z_q\|_2 \downarrow$ | Rel. ↑ |
|---|---|---|---|
| Mean | 4.127 | 2.252 | 45.4% |
| Median | 3.906 | 2.261 | 42.1% |
| P90 | 6.087 | 3.207 | 47.3% |

Table 3 reports results on OpenCpop [14] after 500k training steps. FM-Singer reduces MCD and F0 RMSE relative to the cVAE baselines, suggesting that the learned latent transport generalizes across languages and recording conditions. These improvements are consistent with the intended role of CFM: reducing the gap between training-time posterior latents and inference-time prior samples.

Table 4 provides direct evidence that the proposed latent transport reduces the gap between condition-derived latent samples and posterior latent samples. A smaller distance after refinement indicates that FM-Singer makes inference-time latent representations more consistent with those used during training. Although this result does not by itself establish strict causality, it supports our interpretation that latent mismatch reduction is a plausible mechanism behind the observed quality improvement. In addition to the mean and median distances, the reduced P90 value suggests that the proposed refinement also decreases relatively large-distance cases in the upper tail of the distribution. Positive Rel. values further show that the refined latent representations are consistently closer to the posterior latent samples than the original condition-derived latent samples.

Table 5 shows that the proposed method introduces only limited computational overhead while remaining practically efficient for synthesis. Although FM-Singer adds an additional latent refinement step, its runtime remains close to that of the cVAE baseline under the same decoding configuration. More broadly, this result is consistent with the practical advantage of keeping refinement lightweight, since heavily iterative

**TABLE 5. Comparison of the synthesis speed. Speed of n kHz means that the model can generate n×1000 raw audio samples per second. Real-time means the synthesis speed over real-time.**

| | CPU | | GPU | |
|---|---|---|---|---|
| Model | Speed(kHz) | Real-time | Speed(kHz) | Real-time |
| VISinger2 [3] | 91.42 | × 2.07 | 686.05 | × 15.56 |
| VISinger2 NF | 89.37 | × 2.03 | 593.54 | × 13.46 |
| FM-Singer (ours) | 85.83 | × 1.95 | 660.18 | × 14.97 |

generative schemes, including diffusion-based approaches, generally incur higher runtime cost.

### F. Qualitative Analysis

To visualize the effect of latent refinement on time–frequency structure and pitch trajectories, Fig. 2 compares mel-spectrograms and pitch contours generated by different systems. Compared with the baseline, the proposed method better preserves fine temporal expressive variations, including vibrato-like oscillatory patterns, while maintaining more stable local spectral structure. This qualitative tendency is consistent with the improvements observed in the objective and perceptual evaluations. In contrast, FM-Singer produces mel patterns with clearer harmonic structures and pitch trajectories that more closely follow the reference, consistent with the goal of injecting posterior-like expressive cues into inference-time latents through learned transport.

### G. Discussion

The main benefit of the proposed method is that it addresses the mismatch at its origin. During training, the decoder is optimized using latent representations inferred from the target singing signal, whereas at inference it must rely on latent representations predicted only from the conditioning inputs. Because the decoder is exposed to different latent conditions during training and synthesis, fine expressive acoustic details can be weakened at inference. By refining condition-derived latent representations toward posterior-like latent regions, FM-Singer helps reduce this discrepancy and provides the decoder with inputs that are more consistent with its training condition. Since the refinement operates in latent space, it remains lightweight and integrates naturally with an efficient cVAE backbone without requiring heavy iterative refinement at high resolution.

This interpretation is also consistent with the qualitative and quantitative results. Reducing latent mismatch does not directly alter the predicted F0 trajectory, but it can improve the fidelity of expressive acoustic realization, including vibrato-like modulation, micro-variations, and subtle timbral detail. In this sense, the proposed refinement step can be viewed as a lightweight way to improve expressiveness in cVAE-based singing voice synthesis without redesigning the backbone decoder. More broadly, these results suggest that making inference-time latent representations more consistent with training-time latent conditions is a useful direction for improving practical cVAE-based SVS systems.

## V. Conclusion

This paper presented FM-Singer, a flow-matching-based latent refinement framework for cVAE-based singing voice synthesis. Rather than redesigning the acoustic decoder, the proposed method focuses on reducing the discrepancy between condition-derived latent representations used at inference and target-aware latent representations observed during training. To this end, FM-Singer learns a continuous vector field that transports inference-time latent samples toward posterior-like latent regions and refines them through ODE-based integration before waveform generation. Experimental results on Korean and Chinese benchmarks indicate that this latent refinement improves objective metrics and perceptual quality while maintaining the efficiency of a strong parallel synthesis backbone. Future work includes exploring alternative probability paths beyond linear interpolation, incorporating more explicit technique or style conditioning into the vector field, and reducing integration cost through distillation or other low-step approximations. More broadly, these results suggest that reducing training-inference latent mismatch is a useful direction for improving practical cVAE-based SVS systems.

## Acknowledgment

This work was supported by the Ministry of Science and ICT (MSIT), Korea, and the National IT Industry Promotion Agency (NIPA) through the Advanced GPU Utilization Support Program. Model training was conducted using GPU resources purchased through the “Convergence Open Shared System” Project, supported by the Ministry of Education and the National Research Foundation (NRF) of Korea.